\documentclass[aps,prd,nofootinbib,onecolumn,a4paper]{revtex4}
\usepackage{bm,latexsym,amsmath,amssymb,amsfonts,mathrsfs}
\usepackage[dvipdfmx]{graphicx}
\usepackage{ascmac}
\usepackage{physics}
\usepackage{color,float}

\newcommand{\simgt}{\lower.5ex\hbox{$\; \buildrel > \over \sim \;$}}
\newcommand{\simlt}{\lower.5ex\hbox{$\; \buildrel < \over \sim \;$}}

\begin{document}

\title{\large Effective description of a suspended mirror coupled to cavity light\\
--{Limitations of Q-enhancement due to normal-mode splitting by an optical spring}-- }%

\author{Yuuki Sugiyama,$^{1}$ Tomoya Shichijo,$^{1}$ Nobuyuki Matsumoto,$^{2}$ Akira Matsumura,$^{1}$  Daisuke Miki,$^{1}$   Kazuhiro Yamamoto,$^{1,3}$}

\affiliation{$^1$Department of Physics, Kyushu University, 744 Motooka, Nishi-Ku, Fukuoka 819-0395, Japan}

\email{sugiyama.yuki@phys.kyushu-u.ac.jp}

\affiliation{$^2$Department of Physics, Faculty of Science, Gakushuin University, 1-5-1, Mejiro, Toshima, Tokyo, 171-8588 Japan}

\affiliation{
$^3$Research Center for Advanced Particle Physics, Kyushu University, 744 Motooka, Nishi-ku, Fukuoka 819-0395, Japan
}

\begin{abstract}
Pendulums have long been used as force sensors due to their ultimately low dissipation (high-quality factor) characteristic. 
They are widely used in the measurement of the gravitational constant, detection of gravitational waves, and determination of ultralight dark matter. 
Furthermore, it is expected that the quantum nature of gravity will be demonstrated by performing quantum control for macroscopic pendulums.  
Recently, we have demonstrated that quantum entanglement between two 
pendulums can be generated using an optical spring [D. Miki, N. Matsumoto, A. Matsumura, T. Shichijo, Y. Sugiyama, K. Yamamoto, and N. Yamamoto, arXiv:2210.13169 (2022)]; however, we have ignored that an optical spring can reduce the quality factor (Q-factor) by applying normal-mode splitting between the pendulum and rotational modes possessing relatively high dissipation. 
Herein, we analyze a system composed of a cylinder suspended using a beam (a suspended mirror, i.e., a pendulum) and an optical spring to consider normal-mode splitting. 
The reduction in Q-factor is determined only by the beam parameters: the ratio of the radius of the mirror to the length of the beam, and the ratio of the frequency of the rotational mode to the pendulum mode in the absence of cavity photons.
In our analysis, we find that the reduction factor $4.38$ is reproduced, which is consistent with the experimental result in Matsumoto \textit{et al.} [N. Matsumoto, S. B. Catan$\tilde{\text{o}}$-Lopez, M. Sugawara, S. Suzuki, N. Abe, K. Komori, Y. Michimura, Y. Aso, and K. Edamatsu, Phys. Rev. Lett. 122, 071101 (2019)]. 
Our analysis shows that low dissipation (high quality) can be reached using an optical spring for the realistic pendulum system considering the rotational degree of freedom.
\end{abstract}

\maketitle

\section{Introduction}
Pendulums are sensitive devices considered for force sensing. 
As a pendulum can exhibit ultimately low dissipation, thermal fluctuating forces acting on the pendulum decrease, such that it can be used as a noise-less probe in a force sensor. 
Pendulums are used in various experiments for the measurement of the gravitational constant~\cite{Parks}, direct detection of gravitational waves \cite{Abbott}, and determination of ultralight dark matter \cite{Vermeulen,Graham,Carneyd}. 
To determine external forces acting on a pendulum, an optical cavity has been used \cite{Aspelmeyer, Chen}, which provides an effective way to realize a macroscopic object in the quantum state with continuous measurement and feedback control \cite{MY}. 
When a high reflectivity mirror is suspended as a pendulum and is set at the end of an optical cavity, small displacements of the pendulum mode can be observed as large signals in the output light phase and amplitude. 
Furthermore, an optical cavity affects the mechanical response of a pendulum via the radiation pressure of light generated by an optical spring. 

The optical spring effect is prominent in cavity optomechanics, which was first discussed by Braginsky et al~\cite{Braginsky1, Braginsky2}.
When an optical cavity with a suspended mirror is detuned from the resonance on an input laser, intracavity power linearly depends on the displacement of the mirror. 
As the mirror experiences the change in momentum by reflecting the light, optical restoring forces occur in the detuned cavity, which is known as the optical spring. 
Although damping forces occur due to finite light speed, they do not enhance thermal fluctuating forces on the mirror. 
This is because the optical frequency is so high that thermal photon occupation becomes almost zero. Thus, the optical spring effect can effectively enhance the quality factor beyond the conventional material limit~\cite{Corbitt}. 
However, the quality factor reduces when the rotation of a pendulum changes the cavity length so that the rotational mode couples with the pendulum mode. 
When two modes exchange energy between themselves faster than the time taken for dissipation, the coupled oscillator exhibits normal-mode splitting, which is the main feature of strong coupling~\cite{Novotny}. 
As the rotational mode demonstrates larger dissipation than the pendulum mode, the enhancement of the quality factor is limited, as reported in Ref.~\cite{Kimble1} for a small pendulum trapped in an optical standing wave. 
With regard to the optical spring effect, the limitation of Q-enhancement caused by normal-mode splitting has not been theoretically studied yet. 

In this paper, we consider the optomechanical system consisting of an external input laser (photon) and an optical cavity with a mirror suspended by a beam to include the degree of freedom of a rotational (pitch) mode. 
Although previous studies on radiation pressure applied to the rotational mode~\cite{Solimeno, Sidles, Matsumoto3} exist, they are not on an optical spring system, including pendulum and rotational modes. 
We show that the enhancement of the Q-factor using an optical spring is decreased by a factor of $4.38$ in our model due to normal-mode splitting, which is consistent with the experiment~\cite{Matsumoto2}. 
We discuss that the quantum control of the pendulum is possible when the mechanical dissipation of the pendulum inversely depends on frequency (structural damping model) considering the reduction factor. 
This study is considered to be one of the milestones toward achieving more accurate control of the macroscopic pendulum in quantum states ~\cite{Schmole,Genes,Lopez,Matsumoto2,Michimura,Miki3,Shichijo}.
We expect that optically trapped macroscopic mirrors will significantly contribute to the ultimate force sensing and observation of the quantum nature of gravity in optomechanical systems~\cite{Blaushi,Krisnanda,Miao,Datta,Matsumura,Miki2} in association with Refs.~\cite{Bose,Marletto} 
and stimulated works~\cite{Miki,Carney,Nongaussian,Sugiyama1,Belenchia,Danielson,Sugiyama2,LG}.

This paper is organized as follows: Sec. II introduces the beam model coupled to cavity light. 
We derive equations of motion, which are solved using a perturbative method. 
In Sec. III, we investigate the behavior of background solutions in the steady-state. 
In Sec. IV, we analyze the perturbative equations. 
We demonstrate that the perturbative equations yield an effective theory of the pendulum and rotational modes with mode mixing in the low-frequency region, which enables us to calculate the quality factor with respect to these modes. 
Sec. V concludes the paper. 
The Appendix provides a detailed solution for the steady-state.
\section{Beam model}
\def\x{{q}}
\def\A{{x}}
\def\B{{y}}

\begin{figure}[b]
  \centering
  \includegraphics[width=0.7\linewidth]{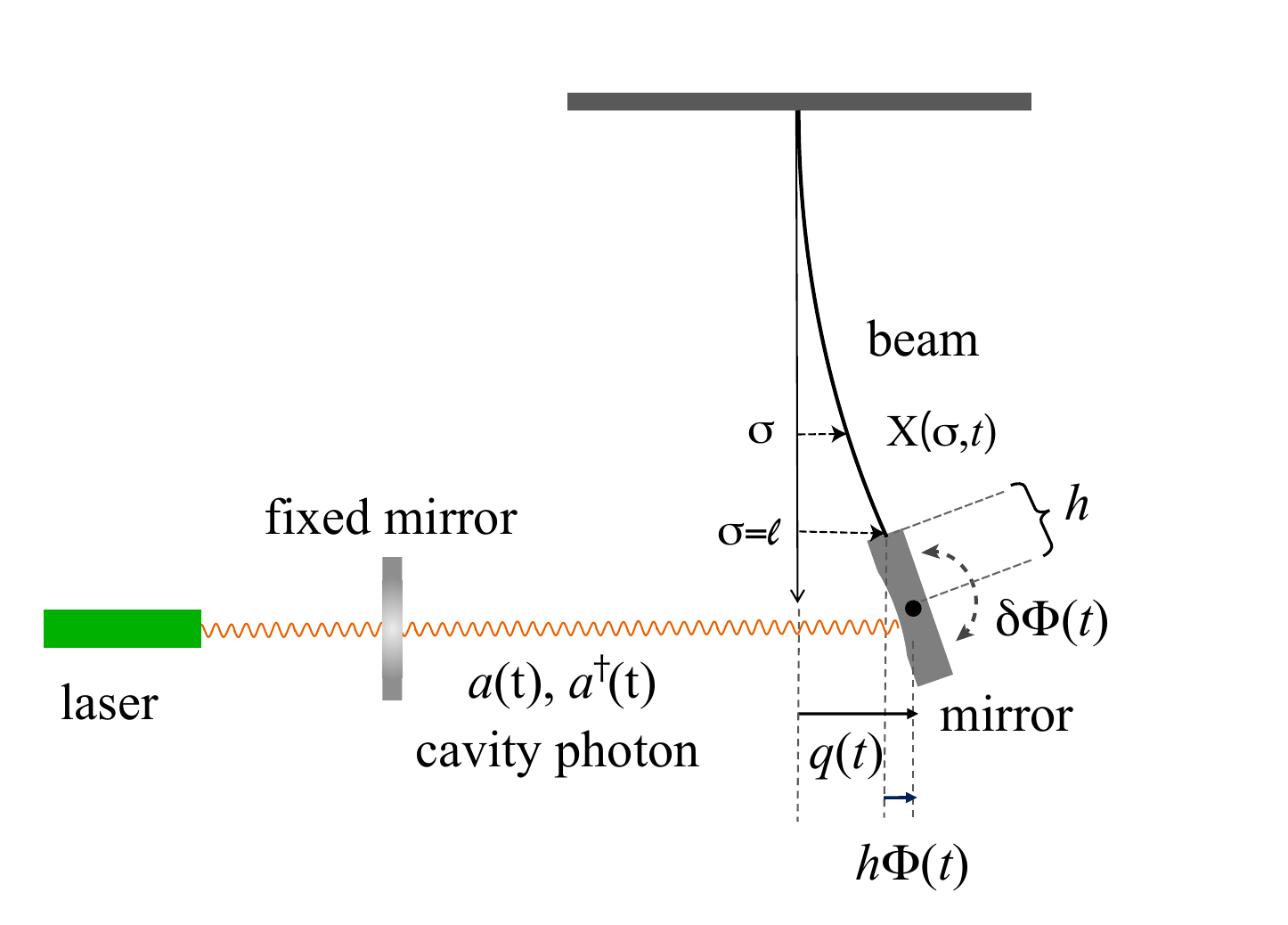}
  \caption{Schematic of the beam model connected to cavity light.}
  \label{fig:beamf1}
\end{figure}

In this section, we review the beam model discussed in Ref.~\cite{Saulson}, and consider an extended model combined with the optical cavity mode.
The beam model is composed of an elastic beam with Young's modulus $E$, area moment of inertia $I$, total length $\ell$, and density of $\rho$ per unit length.
As shown in Figure~\ref{fig:beamf1}, the physical degrees of freedom of the beam model are described using the transverse vibration modes $X(t, \sigma)$ ($0\leq \sigma \leq \ell$).
One end of the beam is fixed, which is described by the boundary conditions 
\begin{eqnarray}
X(t,0)=0, ~~~~~~\frac{\partial X(t, \sigma)}{\partial \sigma}\biggr| _{\sigma=0}=0.
\label{bc1}
\end{eqnarray}
The position of the center of mass with respect to the mirror is described by 
\begin{eqnarray}
\x(t)=X(t, \sigma=\ell)+h\Phi(t)
\label{bc2}
\end{eqnarray}
considering the distance $h$ between the center of mass (mirror radius) and the end of the beam and 
\begin{eqnarray}
\Phi(t)=\frac{\partial X(t, \sigma)}{\partial \sigma}\biggr |_{\sigma=\ell}.
\label{bc3}
\end{eqnarray}
The mirror is assumed as a cylinder, and the moment of inertia $J$ around the axis of rotation is defined by $J=M(3h^2+D^2)/12$ with mass $M$ and thickness $D$.

Under these assumptions, the action of the beam model is described by
\begin{align}
S=\int dt \big(K-V\big)
\end{align}
with the kinetic energies of the beam and the mirror
\begin{align}
K=\frac{1}{2}\int_{0}^{\ell}d\sigma \rho \Big(\frac{{\partial X}}{{\partial t}}\Big)^2+\frac{1}{2}M\Big(\frac{{d{\color{red}q}}}{dt}\Big)^2+\frac{1}{2}J\Big(\frac{{d\Phi}}{dt}\Big)^2,
\end{align}
and the potential energies
\begin{align}
V=\frac{1}{2}\int_{0}^{\ell}d\sigma T\Big(\frac{{\partial X}}{{\partial \sigma}}\Big)^2+\frac{1}{2}Th\Phi^2+\frac{1}{2}\int_{0}^{\ell}d\sigma EI\Big(\frac{{\partial^2X}}{{\partial \sigma^2}}\Big)^2,
\end{align}
where $T=Mg$ denotes the tension of the beam and $g$ denotes the gravitational acceleration.
The last term of the potential energy denotes the elastic energy.
From the action, we derive the following equations of motion as
\begin{align}
\rho \ddot{X}&=TX^{\prime \prime}-EI\frac{\partial^4X}{\partial \sigma^4},
\\
\quad
M\ddot{\x}&=-T\Phi+EI\left.\frac{\partial^3X}{\partial \sigma^3}\right|_{\sigma=\ell},
\label{mqdd}
\\
\quad
J\ddot{\Phi}&=-hEI\left.\frac{\partial^3X}{\partial \sigma^3}\right|_{\sigma=\ell}-EI\left.\frac{\partial^2X}{\partial \sigma^2}\right|_{\sigma=\ell}.
\end{align}
Then, we further include the cavity photon coupled to the mirror, whose Hamiltonian is defined as 
\begin{align}
    H
    &=\hbar\omega_{c}a^{\dagger}a-\hbar\frac{\omega_{c}}{L}\x a^{\dagger}a
    +i\hbar {\cal E}(a^{\dagger}e^{-i\omega_{L}t}-ae^{i\omega_{L}t}),
    \label{hamiltonian}
\end{align}
where $a (a^{\dagger})$ is the annihilation (creation) operator of the cavity photon field with the cavity frequency $\omega_{c}$, and $L$ is the cavity length.
The last term denotes the input laser with frequency $\omega_{L}$ and the amplitude $|{\cal E}|=\sqrt{2P\kappa/\hbar\omega_{L}}$, where $P$ is the input laser power and $\kappa$ is the optical decay rate. 
Herein, we assume $\omega_{\text{L}} \simeq \omega_{\text{c}}$.
The equation of motion of the cavity photon is as follows:
\begin{align}
\dot{a}&=-(\kappa+i(\omega_{\text{c}}-\omega_0))a+iG_{0}a\x+\mathcal{E},
\end{align}
and Eq.~\eqref{mqdd} is modified as
\begin{align}
M\ddot{\x}&=-T\Phi+EI\left.\frac{\partial^3X}{\partial \sigma^3}\right|_{\sigma=\ell}+\hbar G_{0}a^{\dagger}a,
\label{mirroreq}
\end{align}
where we defined $G_{0}=\omega_{c}/L$. 

We solve these equations by considering the perturbations of the system around a steady-state. 
Then, we decompose and combine the variables for the steady-state background, which is expected, and the fluctuation part is defined as follows:
\begin{align}
X=\bar{X}+\delta X,\ \x=\bar{\x}+\delta \x,\ \Phi=\bar{\Phi}+\delta \Phi,\ a=\bar{a}+\delta a.
\end{align}
Assuming the steady-state background, we obtain
\begin{align}
\dot{\bar{X}}=0,\ \dot{\bar{\x}}=0,\ \dot{\bar{\Phi}}=0,\ \dot{\bar{a}}=0,
\label{perturbation}
\end{align}
which helps yield
\begin{align}
0&=T\bar{X}^{\prime \prime}-EI\frac{\partial^4\bar{X}}{\partial \sigma^4},
\\
0&=-T\bar{\Phi}+EI\left.\frac{\partial^3\bar{X}}{\partial \sigma^3}\right|_{\sigma=\ell}+\hbar G_{0}|\bar{a}|^2,
\\
0&=-hEI\left.\frac{\partial^3\bar{X}}{\partial \sigma^3}\right|_{\sigma=\ell}-EI\left.\frac{\partial^2\bar{X}}{\partial \sigma^2}\right|_{\sigma=\ell},
\\
0&=-(\kappa+i(\omega_{\text{c}}-\omega_{\text{L}}-G_{0}\bar{\x}))\bar{a}+\mathcal{E}.
\end{align}
The perturbation equations are derived as
\begin{align}
\rho \delta\ddot{X}&=T\delta X^{\prime \prime}-EI\frac{\partial^4 \delta X}{\partial \sigma^4},
\\
M\delta \ddot{\x}&=-T\delta\Phi+EI\left.\frac{\partial^3\delta X}{\partial \sigma^3}\right|_{\sigma=\ell}+\hbar G_{0}(\bar{a}^{*}\delta a+\bar{a}\delta a^{\dagger}),
\label{radforce}
\\
J\delta \ddot{\Phi}&=-hEI\left.\frac{\partial^3\delta X}{\partial \sigma^3}\right|_{\sigma=\ell}-EI\left.\frac{\partial^2\delta X}{\partial \sigma^2}\right|_{\sigma=\ell},
\\
\delta \dot{a}&=-(\kappa+i(\omega_{\text{c}}-\omega_{\text{L}}))\delta a+iG_{0}(\bar{\x}\delta a+\bar{a}\delta \x).
\end{align}
Herein, we consider the phase reference of the cavity field $\bar{a}=\alpha_{\text{s}}e^{i\theta}$, i.e., $\alpha_{\text{s}}$ is real and positive.
Thus, we rewrite the cavity photon equation as
\begin{align}
\delta \dot{a}^{\prime}&=-(\kappa+i(\omega_{\text{c}}-\omega_{\text{L}}-G_{0}\bar{\x}))\delta a^{\prime}+iG_{0}\alpha_{\text{s}}\delta \x,
\label{ddap}
\end{align}
where we have defined $\delta a^{\prime} = e^{-i\theta}\delta a$.
The last term of Eq.\eqref{radforce} denotes the radiation-pressure force $F_{\text{opt}}$, 
which is defined in the frequency domain as
\begin{align}
\tilde{F}_{\text{opt}}(\omega)
:=
\hbar G_{0} (\bar{a}^{*}\delta \tilde{a}(\omega)+\bar{a}\delta \tilde{a}^{\dagger}(-\omega))
=
\frac{\hbar G^2\Delta}{(\kappa-i\omega)^2+\Delta^2}\delta \tilde{\x}(\omega),
\end{align}
where the quantities with the tilde symbols represent those of the Fourier expansion.
Eq.~\eqref{ddap} yields
\begin{align}
-i\omega \delta \tilde{a}^{\prime}(\omega)
=
-(\kappa+i\Delta)\delta \tilde{a}^{\prime}(\omega)+i\frac{G}{\sqrt{2}}\delta \tilde{\x}(\omega)
\end{align}
considering effective optical detuning $\Delta$ and effective optomechanical coupling $G$ defined by
\begin{align}
\Delta =\Delta_{0}-G_{0}\bar{\x},
\quad
\Delta_{0}=\omega_{\text{c}}-\omega_{\text{L}},
\quad
G=\sqrt{2}G_{0}\alpha_{\text{s}}.
\end{align}
Hence, we obtain the optical spring constant by differentiating $\tilde{F}_{\text{opt}}(\omega)$ with respect to $\delta \tilde{\x}(\omega)$ as $k_{\text{opt}}
:=-{d \tilde{F}_{\text{opt}}}/{d \delta \tilde{\x}(\omega)}$.
Then, we introduce the amplitude quadrature $\delta \A$ and phase quadrature $\delta \B$ and transform the cavity field as
\begin{align}
\sqrt{2}\delta \A\equiv \delta a^{\prime}+\delta {a^{\prime}}^{\dagger},
\quad 
\sqrt{2}i \delta \B\equiv\delta a^{\prime}-\delta {a^{\prime}}^{\dagger}.
\end{align}
As a result, the fluctuation equation of the cavity field is defined as
\begin{align}
\delta \dot{\A}=-\kappa \delta \A+\Delta \delta \B,
\quad
\delta \dot{\B}=-\kappa \delta \B-\Delta \delta \A+G\delta \x,
\end{align}
In the next two sections, we consider the behavior of the steady-state solution and fluctuation solutions.

\section{steady-state solutions}
\begin{figure}[H]
  \centering
  \includegraphics[width=0.5\linewidth]{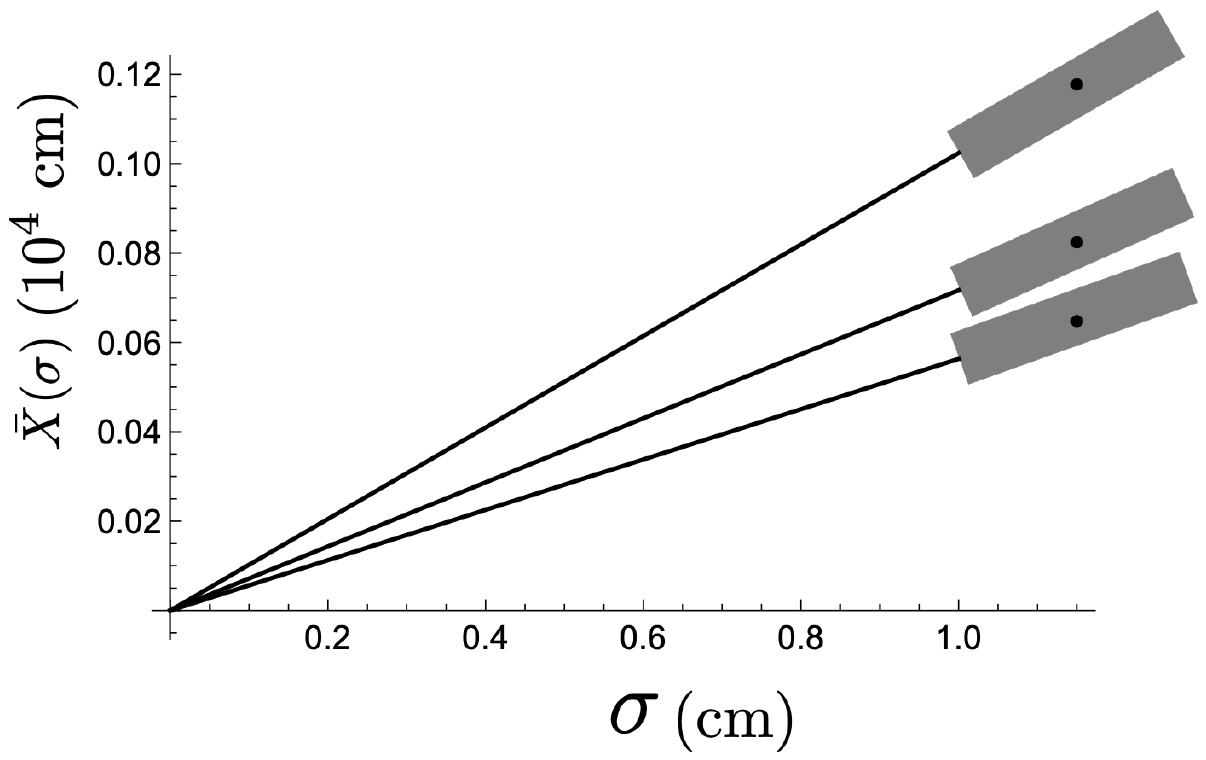}
  \caption{Behavior of the background solution.
  The black lines denote the functions of laser powers $P=0.11 \mathrm{~mW}$, $P=0.16 \mathrm{~mW}$, and $P=0.20 \mathrm{~mW}$ from bottom to top.
  The amplitude is enhanced by $10^4$ times.}
  \label{fig:Background}
\end{figure}

We define the steady-state solution of the beam $\bar{X}(\sigma)$ as
\begin{align}
\bar{X}(\sigma)
&=
\frac{\hbar G_{0}\alpha^2_{\text{s}}}{T}\tilde{X}(\sigma)
\approx \frac{\hbar G_{0}\alpha^2_{\text{s}}}{T}\sigma,
\label{BX}
\end{align}
where $\tilde{X}(\sigma)$ is defined by
\begin{align}
\tilde{X}(\sigma)
&=
\sigma
+
\sqrt{\frac{EI}{T}}\frac{-1+h\sqrt{T/EI}}{(1+h\sqrt{T/EI})e^{2\ell\sqrt{T/EI}}+1-h\sqrt{T/EI}}\big(e^{\sqrt{T/EI}\sigma}-1\big)
\nonumber
\\
&
\quad
+\sqrt{\frac{EI}{T}}e^{2\ell\sqrt{T/EI}}\frac{1+h\sqrt{T/EI}}{(1+h\sqrt{T/EI})e^{2\ell\sqrt{T/EI}}+1-h\sqrt{T/EI}}\big(e^{-\sqrt{T/EI}\sigma}-1\big),
\end{align}
and used the approximations $\ell\sqrt{T/EI} \gg 1$ and $h\sqrt{T/EI} \gg 1$ in the last equality of Eq.~\eqref{BX}.
Thus, the beam's background solution defines a straight line inclined with the coefficient $\hbar G_{0}\alpha^2_{\text{s}}/T$.
Moreover, the position of the mirror is determined by $\bar{X}(\sigma)$ as
\begin{align}
\bar{\x}&=\bar{X}(\ell)+h\left.\frac{\partial}{\partial \sigma}\bar{X}\right|_{\sigma=\ell}
\approx
\frac{\hbar G_{0}\alpha^2_{\text{s}}}{T}(\ell+h).
\end{align}
The solution of the cavity field is defined as
\begin{align}
\bar{a}=\alpha_{\text{s}}e^{i\theta}=\frac{\mathcal{E}}{\kappa+i\Delta},
\end{align}
and this result leads to
\begin{align}
\alpha^2_{\text{s}}
=
\frac{|\mathcal{E}|^2}{\kappa^2+\Delta^2}
=
\frac{|\mathcal{E}|^2}{\kappa^2+(\Delta_{0}-G_{0}\mathcal{G}\alpha^2_{\text{s}})^2},
\label{alphas}
\end{align}
where we defined $\mathcal{G}=\hbar G_{0}(\ell+h)/T$.
This is a cubic algebraic equation with respect to $\alpha^2_{\text{s}}$, and exact solutions are given in Appendix \ref{backgroundsol}.
Detuning $\Delta$ can be adjusted experimentally by varying the cavity length $L$, i.e., the cavity frequency $\omega_{\text{c}}$ within which the approximation of $\omega_{\text{c}}\sim\omega_{\text{L}}$ holds.
The optical spring frequency $\Omega_0$ can be varied by maintaining the laser power $P$ and laser frequency $\omega_{\text{L}}$ constant and changing $\Delta$.
However, we consider that $\Delta=\Delta_{0}-G_{0}\mathcal{G}\alpha^2_{\text{s}}$ is a constant, which can always be achieved by adjusting the laser frequency $\omega_{\text{L}}$ (e.g., see Ref. \cite{Vittorio}).
Then, we introduce the optical enhanced spring frequency $\Omega_{0}=2\pi f \mathrm{~(rad/s)}$ as
\begin{align}
\Omega_{0}
=
2\pi f
&:=
\sqrt{\frac{\Re[k_{\text{opt}}]}{M}},
\end{align}
where considering the real part of the optical spring constant $k_{\text{opt}}$ is justified by assuming the following adiabatic condition: $\kappa \gg \omega$.
Thus, the optical spring frequency becomes
\begin{align}
2\pi f
=
\sqrt{-\frac{1}{M}\frac{\hbar G^2 \Delta}{\kappa^2+\Delta^2}}.
\label{osf}
\end{align}
Introducing the normalized detuning $\delta_{\kappa}:=\Delta/\kappa$ leads to the optical spring frequency, i.e., 
\begin{align}
2\pi f
&=
\sqrt{-\frac{2\hbar G^2_{0}\alpha^2_{\text{s}}}{M\kappa}\frac{\delta_{\kappa}}{1+\delta^2_{\kappa}}}
=
\sqrt{-\frac{2\hbar G^2_{0}\alpha^2_{\text{s}}}{M\kappa}\frac{\delta_{\kappa}}{1+\delta^2_{\kappa}}}
=
\sqrt{-\frac{4\omega_{\text{L}}P}{M\kappa^2}\Big(\frac{G_{0}}{\omega_{\text{L}}}\Big)^2\frac{\delta_{\kappa}}{(1+\delta^2_{\kappa})^2}},
\end{align}
and the effective optomechanical coupling constant $G$ is calculated as
\begin{align}
G
=
\sqrt{2}G_{0}\alpha_{\text{s}}
=
\sqrt{2}G_{0}\sqrt{\frac{|\mathcal{E}|^2}{\kappa^2+\Delta^2}}
=2G_{0}\sqrt{\frac{P}{\kappa \hbar \omega_{\text{L}}}\frac{1}{1+\delta^2_{\kappa}}}.
\end{align}
These results imply that the optical spring frequency and the coupling constant vary with the intensity of the incident light of the laser. 
In our analysis, the frequency splitting occurs near the frequency $\Omega_{0} \approx 170 \mathrm{~rad/s}$; thus, we focus on the scenario near frequency anticrossing.
The results are summarized in TABLE \ref{tab:fpG}, where we used the parameters of TABLE \ref{tab:parameter}.
The behavior of the background solution of the beam is shown in Fig.~\ref{fig:Background}.
The results show that the amplitude increases with the input laser power increases.
\begin{table}[H]
  \centering
  \begin{tabular}{|c|c|c|}
  \hline
  $2\pi f$ $\mathrm{~(rad/s)}$& laser power $\mathrm{~(mW)}$& $G\times10^{18}\mathrm{~(cm^{-1} s^{-1})}$\\
  \hline
  150  & 0.11  & 3.79 \\
  170  & 0.14  & 4.27 \\
  200  & 0.20  & 5.11 \\
  \hline
  \end{tabular}
  \caption{Three typical solutions of the background equation.}
    \label{tab:fpG}
\end{table}
\noindent

\begin{table}[t]
  \centering
  \begin{tabular}{|c|c|c|c|}
   \hline
   Symbol & Description & Value & Dimension \\
   \hline
$\kappa/2\pi$ &Optical decay rate & $ 8.2 \times 10^{5}$ & $\mathrm{~Hz}$ \\
$\omega_{\text{c}}/2\pi \sim \omega_{\text{L}}/2\pi$ & Cavity resonance frequency & $ 2.818\times10^{14}$ &  $\mathrm{~Hz}$\\
$\Delta/\kappa=\delta_{\kappa}$ & Normalized detuning & $-0.0584$ & \\
\hline
$M$ & Mirror mass & $7.71 \times 10^{-3}$ & $\mathrm{~g}$ \\
$D$ & Mirror thickness & $0.05$ & $\mathrm{~cm}$ \\
$L$ & Cavity length & 10.00 & $\mathrm{cm}$\\
$\ell$ & Length of beam & 1.00 & $\mathrm{~cm}$\\
$h$ & Mirror radius & 0.15 & $\mathrm{~cm}$ \\
$J$ & Moment of inertia & $4.50 \times 10^{-5}$ & $\mathrm{~g} \mathrm{~cm}^{2}$ \\
$E_{0} I$ & Flexural rigidity & $3.583 \times 10^{-6}$ & $\mathrm{g} \mathrm{~cm}^{3} / \mathrm{s}^{2}$\\
$\phi$ & Internal loss factor & $10^{-3}$ & \\
$\rho$ & Mass per unit length & $1.72 \times 10^{-8}$ & $\mathrm{~g} / \mathrm{cm}$ \\
\hline
$g$ & Gravitational acceleration & 980.00 & $\mathrm{~cm} / \mathrm{s}^{2}$ \\
$c$ & Speed of light & $2.998 \times 10^{10}$ & $\mathrm{~cm} / \mathrm{s}$\\
$\hbar$ & Reduced Planck constant & $1.05\times10^{-27}$ & $\mathrm{~g} \mathrm{~cm^2}/\mathrm{s}$\\
\hline
  \end{tabular}
    \caption{Parameters}
    \label{tab:parameter}
\end{table}

\section{fluctuation solutions}
In this section, we investigate the perturbations equations in the frequency domain, 
which are explicitly written as 
\begin{align}
-\omega^2\rho \delta\tilde{X}&= T\delta \tilde{X}^{\prime \prime}-EI\frac{\partial^4\delta \tilde{X}}{\partial \sigma^4},
\label{eom1}
\\
-M\omega^2\delta \tilde{\x}&=-T\delta\tilde{\Phi}+EI\left.\frac{\partial^3\delta \tilde{X}}{\partial \sigma^3}\right|_{\sigma=\ell}+\hbar G\delta \tilde{\A},
\label{eom2}
\\
-J\omega^2\delta \tilde{\Phi}&=-hEI\left.\frac{\partial^3\delta \tilde{X}}{\partial \sigma^3}\right|_{\sigma=\ell}-EI\left.\frac{\partial^2\delta \tilde{X}}{\partial \sigma^2}\right|_{\sigma=\ell},
\label{eom3}
\\
-i\omega\delta \tilde{\A}&=-\kappa \delta \tilde{\A}+\Delta \delta \tilde{\B},
\label{eom4}
\\
-i\omega \delta \tilde{\B}&=-\kappa \delta \tilde{\B}-\Delta \delta \tilde{\A}+G\delta \tilde{\x}.
\label{eom5}
\end{align}
We focus on the solution of Eq.~\eqref{eom1}, which has the following solution under 
the boundary conditions defined in Eq.~\eqref{bc1}, 
\begin{align}
\delta \tilde{X}(\omega, \sigma)=A(\cos k \sigma-\cosh k_{\text{e}}\sigma)+B (\sin k \sigma -k/k_{\text{e}}\sinh k_{\text{e}}\sigma),
\label{modef}
\end{align}
where $k$ and $k_{\text{e}}$ are defined as
\begin{align}
k=\sqrt{\frac{-T+\sqrt{T^2+4EI\rho \omega^2}}{2EI}}\approx \sqrt{\frac{\rho}{T}}\omega,
\quad
k_{\text{e}}=\sqrt{\frac{T+\sqrt{T^2+4EI\rho \omega^2}}{2EI}}\approx \sqrt{\frac{T}{EI}}.
\end{align}
Herein, the approximate expressions of $k$ and $k_{\rm e}$ are valid  
under the assumption $T^2 \gg 4EI\rho \omega^2$, i.e., $k_{\text{e}} \gg 2 k $, 
which is valid when we consider that the frequency $\omega/2\pi \ll \times 2.4\times10^6~{\rm Hz}$ in our parameter.
From Eqs.~\eqref{eom4} and \eqref{eom5}, we obtain
\begin{align}
\delta \tilde{\A}=\frac{G\Delta}{\Delta^2+(\kappa-i\omega)^2}\delta \tilde{\x},
\quad
\delta \tilde{\B}=\frac{G(\kappa-i\omega)}{\Delta^2+(\kappa-i\omega)^2}\delta \tilde{\x}.
\end{align}
The solution of Eqs.~\eqref{eom2} and \eqref{eom3} can be obtained as well. 
Conditions \eqref{bc2} and \eqref{bc3} lead to
\begin{align}
{\sqrt{\frac{M}{J}}}
\left[
A(\cos k \ell-\cosh k_{\text{e}}\ell)+B (\sin k \ell -k/k_{\text{e}}\sinh k_{\text{e}}\ell)
\right]
=\sqrt{\frac{M}{J}} \left.\delta \tilde{X}\right|_{\sigma=\ell}
=\sqrt{\frac{M}{J}} \delta \tilde{\x}-h\sqrt{\frac{M}{J}}\delta \tilde{\Phi},
\end{align}
and
\begin{align}
A(-k\sin k\ell-k_{\text{e}}\sinh k_{\text{e}}\ell)+B(k\cos k\ell-k\cosh k_{\text{e}}\ell)=\left.\frac{\partial \delta \tilde{X}}{\partial \sigma}\right|_{\sigma=\ell}=\delta \tilde{\Phi},
\end{align}
where we consider multiplication by $\sqrt{M/J}$ to make the variables dimensionless.
These equations in the matrix form are defined as
\begin{align}
\begin{pmatrix}
\sqrt{\frac{M}{J}}(\cos k \ell-\cosh k_{\text{e}}\ell)
&
\sqrt{\frac{M}{J}}(\sin k \ell -k/k_{\text{e}}\sinh k_{\text{e}}\ell)
\\
-k\sin k\ell-k_{\text{e}}\sinh k_{\text{e}}\ell
&
k\cos k\ell-k\cosh k_{\text{e}}\ell
\end{pmatrix}
\begin{pmatrix}
A\\
B
\end{pmatrix}
=
\begin{pmatrix}
1&-\sqrt{\frac{M}{J}}h\\
0&1
\end{pmatrix}
\begin{pmatrix}
\sqrt{\frac{M}{J}} \delta \tilde{\x}\\
\delta \tilde{\Phi}
\end{pmatrix},
\end{align}
which derives the expressions for the coefficients $A$ and $B$.
\begin{align}
\begin{pmatrix}
A\\
B
\end{pmatrix}
&=C^{-1}
\begin{pmatrix}
\sqrt{\frac{M}{J}} \delta \tilde{\x}\\
\delta \tilde{\Phi}
\end{pmatrix}
=
\begin{pmatrix}
C^{-1}_{00}&C^{-1}_{01}\\
C^{-1}_{10}&C^{-1}_{11}
\end{pmatrix}
\begin{pmatrix}
\sqrt{\frac{M}{J}} \delta \tilde{\x}\\
\delta \tilde{\Phi}
\end{pmatrix}
=
\begin{pmatrix}
C^{-1}_{00} \sqrt{\frac{M}{J}}\delta \tilde{\x}+C^{-1}_{01}\delta \tilde{\Phi}\\
C^{-1}_{10} \sqrt{\frac{M}{J}}\delta \tilde{\x}+C^{-1}_{11}\delta \tilde{\Phi}
\label{coefficient}
\end{pmatrix},
\end{align}
where
\begin{align}
C&=
\begin{pmatrix}
1&\sqrt{\frac{M}{J}}h\\
0&1
\end{pmatrix}
\begin{pmatrix}
\sqrt{\frac{M}{J}}(\cos k \ell-\cosh k_{\text{e}}\ell)
&
\sqrt{\frac{M}{J}}(\sin k \ell -k/k_{\text{e}}\sinh k_{\text{e}}\ell)
\\
-k\sin k\ell-k_{\text{e}}\sinh k_{\text{e}}\ell
&
k\cos k\ell-k\cosh k_{\text{e}}\ell
\end{pmatrix}
\nonumber\\
&
\quad
=
\begin{pmatrix}
\sqrt{\frac{M}{J}}(\cos k \ell-\cosh k_{\text{e}}\ell-h(k\sin k\ell+k_{\text{e}}\sinh k_{\text{e}}\ell))
&
\sqrt{\frac{M}{J}}(\sin k \ell -k/k_{\text{e}}\sinh k_{\text{e}}\ell+hk(\cos k\ell-\cosh k_{\text{e}}\ell))
\\
-k\sin k\ell-k_{\text{e}}\sinh k_{\text{e}}\ell
&
k(\cos k\ell-\cosh k_{\text{e}}\ell)
\end{pmatrix},
\end{align}
and the inverse of the matrix $C$ is defined by
\begin{align}
C^{-1}&=\frac{1}{\det C}
\begin{pmatrix}
k\cos k\ell-k\cosh k_{\text{e}}\ell
&
-\sqrt{\frac{M}{J}}(\sin k \ell -k/k_{\text{e}}\sinh k_{\text{e}}\ell+hk(\cos k\ell-\cosh k_{\text{e}}\ell))
\\
k\sin k\ell+k_{\text{e}}\sinh k_{\text{e}}\ell
&
\sqrt{\frac{M}{J}}(\cos k \ell-\cosh k_{\text{e}}\ell-h(k\sin k\ell+k_{\text{e}}\sinh k_{\text{e}}\ell))
\end{pmatrix}
\end{align}
with
\begin{align}
\det C
=
\sqrt{\frac{M}{J}}
\left[2k+(k_{\text{e}}-k^2/k_{\text{e}})\sin k\ell \sinh k_{\text{e}}\ell -2k\cos k\ell \cosh k_{\text{e}}\ell
\right].
\end{align}
Thus, $A$ and $B$ are written in terms of $\delta \tilde{\x}$ and $\delta \tilde{\Phi}$, which provides the solution for $\delta \tilde{X}(\omega, \sigma)$ 
through Eq.~\eqref{modef}. 
Then, Eqs.~\eqref{eom2} and \eqref{eom3} lead to
\begin{align}
-\omega^2\left(\sqrt{\frac{M}{J}}\delta \tilde{\x}\right)
&=
\sqrt{\frac{M}{J}}
\left[-\frac{T}{M}\delta\tilde{\Phi}+\frac{EI}{M}\left.\frac{\partial^3\delta \tilde{X}}{\partial \sigma^3}\right|_{\sigma=\ell}+\frac{\hbar G}{M}\delta \tilde{\A}
\right]
\nonumber\\
\quad
&=
-\omega^2_{\text{A}}(\omega) \left(\sqrt{\frac{M}{J}}\delta \tilde{\x}\right)+\Delta^2_{\text{A}}(\omega)\delta \tilde{\Phi},
\end{align}
and
\begin{align}
-\omega^2\delta \tilde{\Phi}
&=-\frac{hEI}{J}\left.\frac{\partial^3\delta \tilde{X}}{\partial \sigma^3}\right|_{\sigma=\ell}-\frac{EI}{J}\left.\frac{\partial^2\delta \tilde{X}}{\partial \sigma^2}\right|_{\sigma=\ell}
\nonumber\\
\quad
&=
\omega^2_{\text{B}}(\omega)\Big(\sqrt{\frac{M}{J}}\delta \tilde{\x}\Big)-\Delta^2_{\text{B}}(\omega)\delta \tilde{\Phi}, 
\end{align}
respectively, which are written in the matrix form as 
\begin{align}
-\omega^2
\begin{pmatrix}
\sqrt{\frac{M}{J}} \delta \tilde{\x}\\
\delta \tilde{\Phi}
\end{pmatrix}
=
\begin{pmatrix}
-\omega^2_{\text{A}}(\omega)&\Delta^2_{\text{A}}(\omega)\\
\omega^2_{\text{B}}(\omega)&-\Delta^2_{\text{B}}(\omega)
\end{pmatrix}
\begin{pmatrix}
\sqrt{\frac{M}{J}} \delta \tilde{\x}\\
\delta \tilde{\Phi}
\end{pmatrix}.
\label{befdia}
\end{align}
This represents coupled harmonic oscillators of the two modes: $\delta \tilde{\x}$ (pendulum mode) and $\delta \tilde{\Phi}$ (rotational mode). 
The $\omega$-dependence of the matrix of the right-hand-side of Eq.~\eqref{befdia} does not always allow such a simple interpretation. 
For instance, the $\omega$-dependence of the matrix represents the violin modes reflecting the properties of the beam's degrees of freedom. 
However, we can demonstrate that the  $\omega$-dependence of the matrix disappears by assuming the conditions $T^2 \gg 4EI\rho \omega^2$ and $\kappa \gg \omega$, which are satisfied for the parameters adopted in the present paper (TABLE \ref{tab:parameter}). 
Under the condition, $T^2 \gg 4EI\rho \omega^2$, we can neglect the higher order terms of $\mathcal{O}((1/k_{\text{e}}\ell )^2)$,  $\mathcal{O}((1/k_{\text{e}}h)^2)$, and  $\mathcal{O}(1/k^2_{\text{e}}\ell h)$. 
Thus, we obtain
\begin{align}
\omega^2_{\text{A}}(0)
&\approx
\Omega^2_{0}+\omega^2_{\text{p}}\Big(1+\frac{2}{k_{\text{e}}\ell}\Big),
\label{wa2}
\\
\Delta^2_{\text{A}}(0)
&\approx
\omega_{\text{p}}\omega_{\text{r}}\sqrt{r}\Big[1+\frac{1}{k_{\text{e}}\ell}\Big(2+\frac{1}{r}\Big)\Big],
\\
\omega^2_{\text{B}}(0)
&\approx
\omega_{\text{p}}\omega_{\text{r}}\sqrt{r}\Big[1+\frac{1}{k_{\text{e}}\ell}\Big(2+\frac{1}{r}\Big)\Big],
\\
\Delta^2_{\text{B}}(0)
&\approx
\omega^2_{\text{r}}
\Big[
1+r+\frac{1}{k_{\text{e}}\ell}\Big(2+2r+\frac{1}{r}\Big)
\Big].
\label{db2}
\end{align}
Here, we introduced the parameters of the ratio $r=h/\ell$, and the frequency of the pendulum mode $\omega_{\text{p}}=\sqrt{g/\ell}$ and the rotational mode $\omega_{\text{r}}=\sqrt{Mgh/J}$ in the absence of cavity photons.
Such a system described by Eq.~\eqref{befdia} with Eqs.~\eqref{wa2} $\sim$ \eqref{db2} then reduces to a well-known system of a coupled harmonic oscillator, 
in which mode splitting or resonant conversion occurs \cite{Novotny}. 
We discuss the features of the pendulum mode and rotational mode, and evaluate the quality factor. 

\begin{figure}[H]
  \centering
  \includegraphics[width=0.9\linewidth]{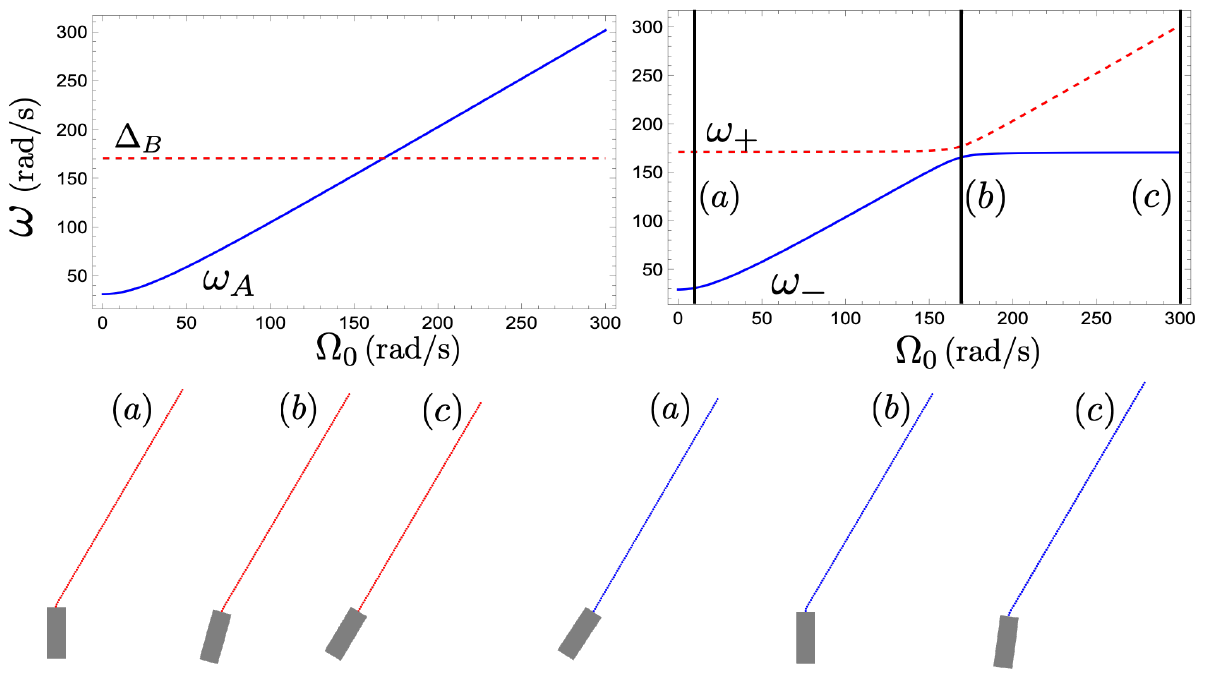}
  \caption{
  Upper panels: Frequency $\omega_A$ and $\Delta_B$ (left panel) and eigenfrequency $\omega_\pm$ (right panel)
  as a function of $\Omega_0=2\pi f$ in the unit of rad/s.
  The left panel represents the frequencies of the diagonal component $\Delta_{\text{B}}(0)$ (red dashed line) and $\omega_{\text{A}}(0)$ (blue solid line).
  The right panel shows the eigenfrequencies $\omega_+$ (red dashed curve) and $\omega_-$ (blue solid curve) due to the mode mixing.  
  Lower panels: Plots of the mode shape of the beam and the mirror at the frequencies (a)~$\Omega_0=10\mathrm{~rad/s}$, (b)~$\Omega_0=170\mathrm{~rad/s}$, and (c)~$\Omega_0=300\mathrm{~rad/s}$, as specified by the black vertical lines in the upper right panel. 
 The left three mode shapes are represented with $\omega_+$, and the right three mode shapes are represented with $\omega_-$, where the arbitrary amplitude is adopted for each mode shape. 
 Here, the assumed parameters are listed in the TABLE II.}
  \label{fig:R}
\end{figure}

The upper left panel of Fig.~\ref{fig:R} plots $\omega_{\text{A}}(0)$ (blue solid curve) and $\Delta_{\text{B}}(0)$ (red dashed curve), the frequencies of the diagonal components of the matrix in Eq.~\eqref{befdia}, as functions of $\Omega_{0}=2\pi f$. 
These are the frequencies of the $\delta q$ and $\delta \Phi$ when the two modes are decoupled. As one can see from Eq.~\eqref{wa2}, $\omega^2_{\text{A}}(0)$ depends on the optical spring frequency $\Omega_{0}=2\pi f$, i.e., the frequency of the pendulum mode is enhanced by the optical spring due to the coupling of the photon. 
On the other hand, $\Delta_{\text{B}}(0)$ does not depend on $\Omega_0$, which characterizes the frequency of the rotational motion.
However, due to the non-diagonal components of the matrix $\Delta_A^2(0)$ and $\omega_B^2(0)$, the mode-mixing appears, which is characterized by the eigenfequencies of the matrix in Eq.~\eqref{befdia}.
The following equation is derived from Eq.~\eqref{befdia} by diagonalizing the matrix
\begin{align}
-\omega^2
\begin{pmatrix}
\delta \tilde{\x}_{\text{d}}\\
\delta \tilde{\Phi}_{\text{d}}
\end{pmatrix}
=-
\begin{pmatrix}
\omega^2_{+}(\Omega_{0})&0\\
0&\omega^2_{-}(\Omega_{0})
\end{pmatrix}
\begin{pmatrix}
\delta \tilde{\x}_{\text{d}}\\
\delta \tilde{\Phi}_{\text{d}}
\end{pmatrix},
\label{eigeneq}
\end{align}
where the eigenfrequencies $\omega_{\pm}$ are introduced by
\begin{align}
\omega^2_{\pm}(\Omega_{0})
&=
\frac{1}{2}
\Big[
\omega^2_{\text{A}}(0)+\Delta^2_{\text{B}}(0)
\pm
\sqrt{
\big(
\omega^2_{\text{A}}(0)-\Delta^2_{\text{B}}(0)
\big)^2
+4\Delta^2_{\text{A}}(0)\omega^2_{\text{B}}(0)}
\Big],
\label{omegapm}
\end{align}
and eigenvectors $\delta \tilde{\x}_{\text{d}}$ and $\delta \tilde{\Phi}_{\text{d}}$ are defined as
\begin{align}
\begin{pmatrix}
\sqrt{\frac{M}{J}}\delta \tilde{\x}\\
\delta \tilde{\Phi}
\end{pmatrix}
&=P
\begin{pmatrix}
\delta \tilde{\x}_{\text{d}}\\
\delta \tilde{\Phi}_{\text{d}}
\end{pmatrix},
\quad
P=
\begin{pmatrix}
\sin \beta&\cos \beta\\
\cos \beta&-\sin \beta
\end{pmatrix}
\label{diago}
\end{align}
with 
$\tan \beta =(\Delta^2_{\text{B}}(0)-\omega^2_{\text{+}})/\omega^2_{\text{B}}(0)=(\omega^2_{\text{-}}-\omega^2_{\text{A}}(0))/\omega^2_{\text{B}}(0)$.
By substituting the eigenvalues $\omega=\omega_{\pm}(\Omega_{0})$ into Eq.~\eqref{eigeneq}, we have $\delta \tilde{\Phi}_{\text{d}}(\omega_{+})=0$ and  $\delta\tilde{\x}_{\text{d}}(\omega_{-})=0$.
Using the variables $\delta \tilde{\x}_{\text{d}}$ and $\delta \tilde{\Phi}_{\text{d}}$, $\delta \tilde{X}(\omega, \sigma)$ is obtained by substituting \eqref{coefficient} and \eqref{diago} into \eqref{modef} as
\begin{align}
\delta \tilde{X}(\omega, \sigma)
&=\Big[\big(C^{-1}_{00}(\omega)\sin \beta +C^{-1}_{01}(\omega)\cos \beta \big)\big(\cos k\sigma-\cosh k_{\text{e}}\sigma\big)
\nonumber\\
&
\quad
+(C^{-1}_{10}(\omega)\sin \beta +C^{-1}_{11}(\omega)\cos \beta \big)\big(\sin k\sigma-k/k_{\text{e}}\sinh k_{\text{e}}\sigma\big)
\Big]\delta \tilde{\x}_{\text{d}}(\omega)
\nonumber\\
&
\quad
+\Big[\big(C^{-1}_{00}(\omega)\cos \beta -C^{-1}_{01}(\omega)\sin \beta \big)\big(\cos k\sigma-\cosh k_{\text{e}}\sigma\big)
\nonumber\\
&
\quad
+(C^{-1}_{10}(\omega)\cos \beta -C^{-1}_{11}(\omega)\sin \beta \big)\big(\sin k\sigma-k/k_{\text{e}}\sinh k_{\text{e}}\sigma\big)
\Big]\delta \tilde{\Phi}_{\text{d}}(\omega).
\end{align}
In particular, at the frequency $\omega=\omega_{\pm}$, $\delta \tilde{X}(\omega, \sigma)$ becomes
\begin{align}
\delta \tilde{X}(\omega_{+}, \sigma)
&=
\Big[\big(C^{-1}_{00}(\omega_{+})\sin \beta +C^{-1}_{01}(\omega_{+})\cos \beta \big)\big(\cos k\sigma-\cosh k_{\text{e}}\sigma\big)
\nonumber\\
&
\quad
+(C^{-1}_{10}(\omega_{+})\sin \beta +C^{-1}_{11}(\omega_{+})\cos \beta \big)\big(\sin k\sigma-k/k_{\text{e}}\sinh k_{\text{e}}\sigma\big)
\Big]\delta \tilde{\x}_{\text{d}}(\omega_{+}),
\label{pmode}
\end{align}
and
\begin{align}
\delta \tilde{X}(\omega_{-}, \sigma)
&=
\Big[\big(C^{-1}_{00}(\omega_{-})\cos \beta -C^{-1}_{01}(\omega_{-})\sin \beta \big)\big(\cos k\sigma-\cosh k_{\text{e}}\sigma\big)
\nonumber\\
&
\quad
+(C^{-1}_{10}(\omega_{-})\cos \beta -C^{-1}_{11}(\omega_{-})\sin \beta \big)\big(\sin k\sigma-k/k_{\text{e}}\sinh k_{\text{e}}\sigma\big)
\Big]\delta \tilde{\Phi}_{\text{d}}(\omega_{-}).
\label{mmode}
\end{align}

Equations.~\eqref{pmode} and \eqref{mmode} are very complicated. 
Furthermore, the numerical approach must be carefully implemented due to the factor $e^{k_{\text{e}}\ell} \sim 5e^{1000}$ when the parameters listed in TABLE \ref{tab:parameter} are adopted. 
Therefore, we expand the coefficients $C^{-1}_{00}(\omega_{\pm})$, $C^{-1}_{01}(\omega_{\pm})$, $C^{-1}_{10}(\omega_{\pm})$, and $C^{-1}_{11}(\omega_{\pm})$ up to the order of $\mathcal{O}{(1/k_{\text{e}}\ell)}$, $\mathcal{O}{(1/k_{\text{e}}h)}$, and $\mathcal{O}{(k/k_{\text{e}})}$.
The upper right panel of Fig.~\ref{fig:R} shows the eigenfrequencies $\omega_{+}$ (red dashed curve) and $\omega_{-}$ (blue solid curve) defined by Eq.~\eqref{omegapm} as a function of the optical spring frequency $\Omega_{0}=2\pi f$.
In the upper right panel of Fig.~\ref{fig:R}, the blue solid curve ($\omega_-$) is regarded as the frequency of the pendulum mode for $\Omega_{0}\simlt170\mathrm{~rad/s}$, while the red dashed curve ($\omega_+$) behaves as the pendulum mode for $\Omega_{0}\simgt170\mathrm{~rad/s}$ because it increases monotonically in proportion to the frequency of the optical spring, where the pendulum mode is well coupled to the photon.
The lower panel of Fig.~\ref{fig:R} shows the mode shape of the beam and the mirror at (a)~$\Omega_0=10\mathrm{~rad/s}$, (b)~$\Omega_0=170\mathrm{~rad/s}$, and (c)~$\Omega_0=300\mathrm{~rad/s}$.
This situation is analogous to normal-mode splitting.
At a low frequency, i.e., $\Omega_0=10\mathrm{~rad/s}$ and high frequency, i.e., $\Omega_0=300\mathrm{~rad/s}$, the two modes are perfectly splitting.
For example, the two-mode shape clearly represents the pendulum and the rotational modes as exhibited by the mode shape labeled with (a) and (c) in the lower panel. 
Therefore, two modes are decoupled.
At around (b) $\Omega_0=170\mathrm{~rad/s}$, the two-mode shape represents a mixture, as shown by the mode shape labeled with (b) in the lower panel.

Then, we introduce the quality factor $Q_{\pm}(\Omega_{0})$, which characterizes the number of times the pendulum oscillates before the damping, 
\begin{align}
Q_{\pm}(\Omega_{0})
:=
\frac{\Re[\omega_{\pm}(\Omega_{0})]}{|\Im[\omega_{\pm}(\Omega_{0})]|},
\end{align}
where $\Im[\omega_{\pm}]$ stems from the structural damping effect by adding the imaginary part into Young's modulus $E$ as $E=E_{0}(1-i\phi(\omega))$~\cite{Saulson}.
Here, $E_{0}$ denotes the real part of Young's modulus, and $\phi(\omega)$ characterizes its imaginary part.
Note that, in reality, $\phi(\omega)$ demonstrates an almost approximated constant with respect to $\omega$ over a large band frequency~\cite{Saulson2}.
To simply compute the quality factor, we expand the imaginary part to the first order of $\phi \ll 1$.

\begin{figure}[H]
  \centering
  \includegraphics[width=0.55\linewidth]{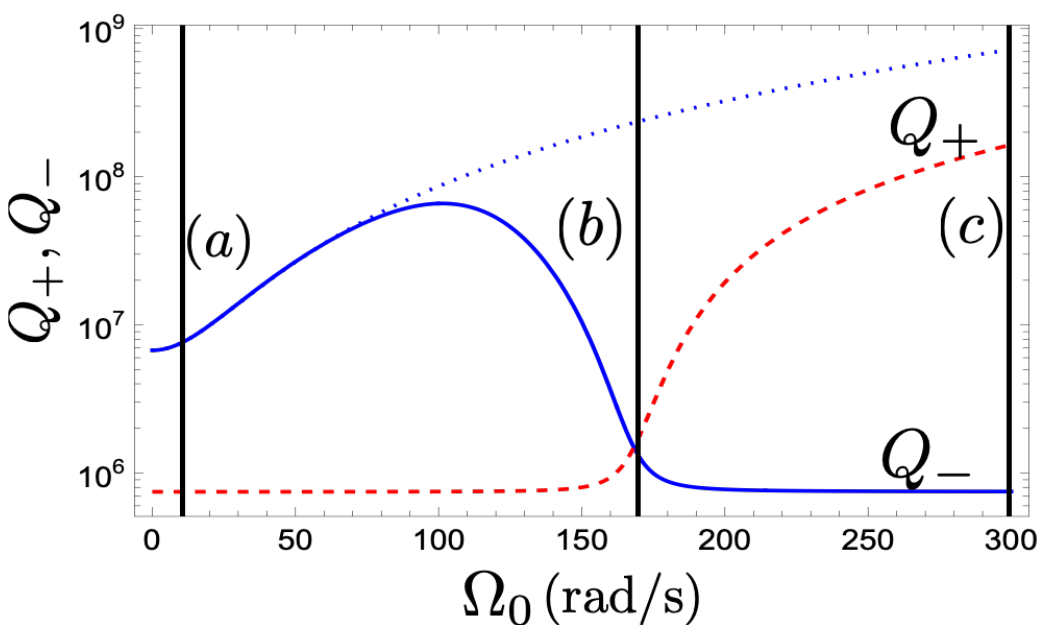}
  \caption{
  Quality factors $Q_+$ (red dashed curve) and $Q_-$ (blue solid curve) as a function of $\Omega_0=2\pi f$ in the unit of rad/s. 
  The blue dotted curve represents the extrapolation of the quality factor from the low frequency.
 Here, the assumed parameters are listed in TABLE II.}
  \label{fig:Q}
\end{figure}
Figure~\ref{fig:Q} plots the quality factor $Q_{+}(\Omega_{0})$ (red dashed curve) and $Q_{-}(\Omega_{0})$ (blue solid curve) as a function of frequency $\Omega_{0}=2\pi f$.
At around $\Omega_{0} \sim 10\mathrm{~rad/s}$, the quality factor of the pendulum mode $Q_{-}(\Omega_{0})$ increases monotonically. 
However, due to the coupling with the rotational mode, the quality factor decreases for $\Omega_0 > 100\mathrm{~rad/s}$.
After mode splitting occurred at $\Omega_{0} \sim 170\mathrm{~rad/s}$, each mode is mixed indistinguishably.
At $\Omega_{0} \sim 300\mathrm{~rad/s}$, the pendulum mode and the rotational mode are perfectly splitting, and $Q_{+}(\Omega_{0})$ behaves as the pendulum mode, which is increased monotonically by the optical spring frequency.
In our model, the quality factor reaches $1.0\times10^8$ at $\Omega_{0} \sim 300\mathrm{~rad/s}$. 
The blue dotted curve in Fig.~\ref{fig:Q} shows the quality factor obtained through extrapolation from the low-frequency behavior.
This curve corresponds to the pendulum mode in the absence of the reduction of the quality factor due to the rotational mode.
Now, we can evaluate how much the quality factor of the pendulum mode is reduced due to the presence of
the rotational mode associated with the mode splitting.
At $\Omega_{0} \sim 300\mathrm{~rad/s}$ where the pendulum mode and the rotational mode are well splitting, the quality factor of the pendulum mode is evaluated by $Q_{+}(\Omega_{0})$, which is represented by the red dashed curve in Fig.~\ref{fig:Q}.
The blue dotted curve in Fig.~\ref{fig:Q}, which is regarded as the pendulum mode in the absence of the rotational mode, is defined by the Taylor expansion of the quality factor of the pendulum mode at low frequency as
\begin{eqnarray}
Q_{-}(\Omega_{0})
\approx 
Q_{-}(0)+\frac{1}{2}
\left(\frac{d^2Q_{-}(\Omega_{0})}{d\Omega^2_{0}}
\right)
\bigg|_{\Omega_{0}=0}\Omega^2_{0}.
\label{extrapolation}
\end{eqnarray}
Thus the reduction of the quality factor due to the coupling of the two modes is given by a function $R(\Omega_{0})$ defined by the ratio of Eq.~\eqref{extrapolation} to $Q_{+}(\Omega_{0})$ as
\begin{align}
R(\Omega_{0})
=
\frac{1}{Q_{+}(\Omega_{0})}
\left(
Q_{-}(0)+\frac{1}{2}
\left(
\frac{d^2Q_{-}(\Omega_{0})}{d\Omega^2_{0}}
\right)
\bigg|_{\Omega_{0}=0}\Omega^2_{0}
\right).
\label{romega}
\end{align}
We note that Eq.~\eqref{romega} is justified in the range $\Omega_{0} > 170\mathrm{~rad/s}$, where the mode splitting occurred.
Using the parameters of Table \ref{tab:parameter}, the quality factor is reduced by $R(300)=4.38$.
This is consistent with a prediction expected from the experimental result of Ref.~\cite{Matsumoto2}.
\begin{figure}[t]
  \centering
  \begin{minipage}[b]{0.48\linewidth}
  \includegraphics[width=1\linewidth]{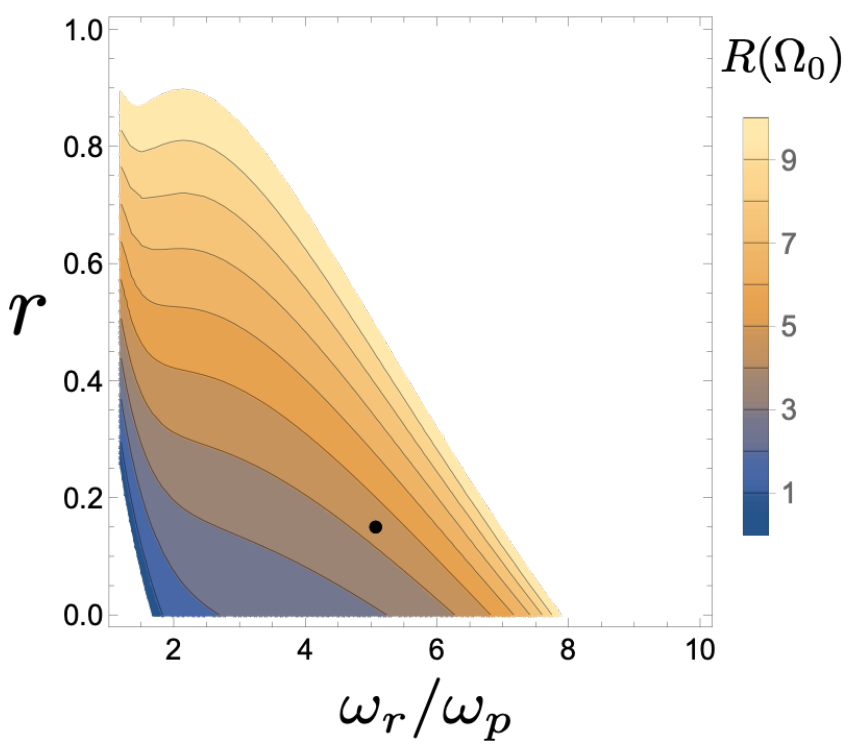}
  \end{minipage}~~~~~~~~~~~~~
  \begin{minipage}[b]{0.50\linewidth}
  \includegraphics[width=1\linewidth]{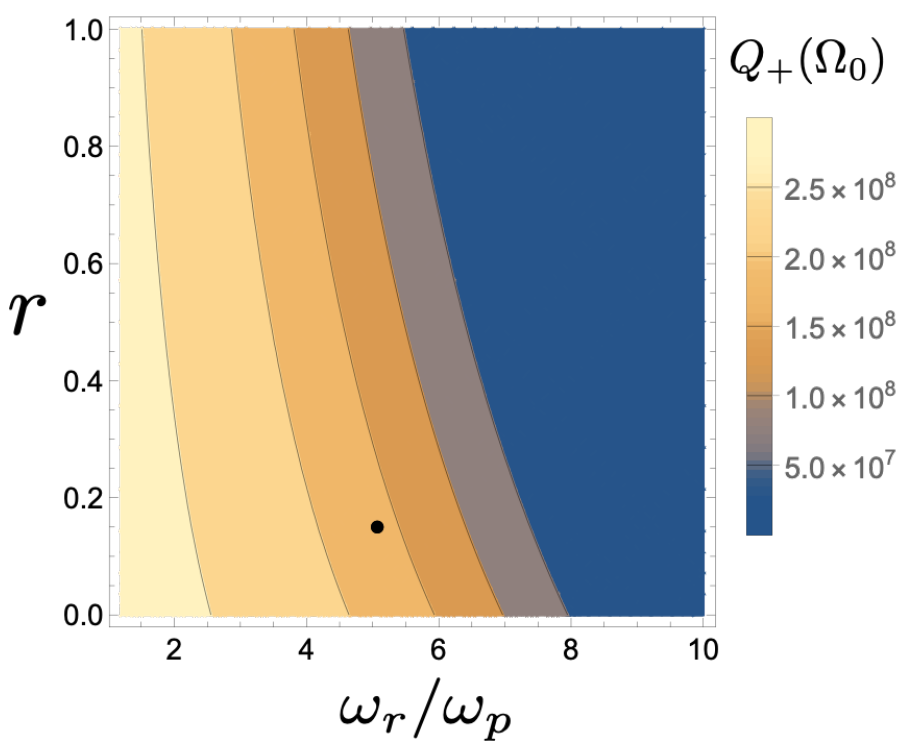}
  \end{minipage}
  \caption{
  The left panel shows $R(\Omega_0)$ defined by Eq.~\eqref{romega}, the ratio of the quality factor extrapolated
  from the low frequency, the right hand side of Eq.~\eqref{extrapolation} 
  to $Q_+(\Omega_0)$. Namely, $R(\Omega_0)$ represents the ratio of the blue dotted curve to the red dashed curve~\eqref{romega} in Fig.~\ref{fig:Q}.
  The right panel plots the quality factor $Q_{+}(\Omega_{0})$ as functions of $\omega_{\text{r}}/\omega_{\text{p}}=\sqrt{Mh\ell/J}$ and $r=h/\ell$.
  Here we assume the ratio $\Omega_{0}/\omega_{\text{p}}$ is fixed as $300/\sqrt{980}$, and, in this parameter, the quality factor $Q_{+}(\Omega_{0})$ behaves as the pendulum mode (see Fig.~\ref{fig:Q}).}
  The black circle covers the parameters of Table \ref{tab:parameter}.
  \label{rp}
\end{figure}

Next, we investigate the parameter dependence of the reduction of the quality factor.
Figure~\ref{rp} shows the behavior of the $R(\Omega_{0})$ and $Q_{+}(\Omega_{0})$ as a function of the normalized rotational frequency $\omega_{\text{r}}/\omega_{\text{p}}$ and $r$, where $\Omega_{0}/\omega_{\text{p}}$ is fixed.
To maintain the quality factor high and reduction low, we need the values of $\omega_{\text{r}}/\omega_{\text{p}}=\sqrt{Mh\ell/J}$ and $r=h/\ell$ to be small at the same time.
The approximate analytical formula is also obtained by considering a large optical spring frequency limit as
\begin{align}
R(\infty)
&=
\lim_{\Omega_{0}/\omega_{\text{p}}\rightarrow\infty}R(\Omega_{0})
\nonumber\\
\quad
&=
\frac{8r^2[-1-r^2(1+r)(\omega_{\text{r}}/\omega_{\text{p}})^6+\lambda+\{-2+\lambda+r^2(\lambda-3)\}(\omega_{\text{r}}/\omega_{\text{p}})^4+\{1+r(2\lambda-3)\}(\omega_{\text{r}}/\omega_{\text{p}})^2]}
{\lambda[-2r\lambda^2+2r\lambda+(1+2r+2r^2)(\omega_{\text{r}}/\omega_{\text{p}})^2\lambda+(1-4r)-(1+r)(\omega_{\text{r}}/\omega_{\text{p}})^4]},
\end{align}
where $\lambda$ is defined by $\lambda=\sqrt{1+2(r-1)(\omega_{\text{r}}/\omega_{\text{p}})^2+(1+r)^2(\omega_{\text{r}}/\omega_{\text{p}})^4}$.
In this limit, the ratio $R(\infty)$ is described by $\omega_{\text{r}}/\omega_{\text{p}}$ and $r$ and is evaluated using our parameters as $R(\infty)\simeq 2.68$.
This result is determined only by the beam model parameters.

\section{conclusion}
We proposed a theoretical model for the detuned optical cavity, including a suspended mirror with a single beam. 
With this model, we consider an optical spring for the pendulum and rotational modes when we ignored all the violin modes, i.e., focusing on the low-frequency vibrations of the beam. 
We succeeded in deriving the analytical solution by decomposing it into a steady-state solution and its fluctuating solution.
The steady-state solution was found in an analytic manner, which behaves in an intuitively correct manner, i.e., the amplitude of the beam increases as the input laser power increases. 
In the perturbative solution, we found that the pendulum and rotational modes form a coupled harmonic oscillator system. 
We demonstrated that the pendulum and rotational modes resonate with each other based on the analogy of a coupled harmonic oscillator, leading to mode mixing of them. 
When mode mixing occurred, two modes cannot be perfectly separated such that the quality factor of the pendulum was reduced by a factor of 4.38. 
This reduction is consistent with the experiment described in Ref.~\cite{Matsumoto2}, in which the quality factor of the pendulum exceeds $1\times10^8$.

The analysis in this paper is based on classical theory.
In the future, it will be necessary to quantize the degrees of freedom of mirrors and beam to understand their quantum behavior, assuming that quantum control will be applied to the proposed model. 
In addition, we aim to develop an experimentally realistic setup.
In the present paper, we ignored the dissipation and fluctuation due to the collision of air molecules and the fluctuation of photons because we focus on the behavior of mode splitting and the evaluation of the Q-factor, but these fluctuation effects are necessary to estimate the magnitude of fluctuation forces~\cite{Shichijo}.

\acknowledgements
We thank N. Yamamoto for discussions related to the topic in the present paper. 
Y.S. was supported by the Kyushu University Innovator Fellowship in Quantum Science.
K.Y. was supported by JSPS KAKENHI, Grant No. 22H05263.
N.M. was supported by JSPS KAKENHI, Grant No. 19H00671 and JST FORESTO, Grant NO. JPMJFR202X.
D.M. is supported by JSPS KAKENHI, Grant No. 22J21267.
\begin{appendix}
\section{Background solution\label{backgroundsol}}
We present the exact solution of equation \eqref{alphas}.
The solutions of Eq. \eqref {alphas} with respect to $\alpha_{\text{s}}$ are described as follows:
\begin{align}
\alpha^2_{\text{s}}
&=
\frac{2 \Delta_{0}}{3 G_{0} \mathcal{G}}
+\frac{\left(\Delta_{0}^{2}-3 \kappa^{2}\right)}{3\mathcal{F}(G_{0}\mathcal{G},\mathcal{E}^2)}
+\frac{\mathcal{F}(G_{0}\mathcal{G},\mathcal{E}^2)}{3\left(G_{0} \mathcal{G}\right)^{2}},
\\
\quad
&
\frac{2 \Delta_{0}}{3 G_{0} \mathcal{G}}
-\frac{(1+i\sqrt{3})}{2}\frac{\left(\Delta_{0}^{2}-3 \kappa^{2}\right)}{3\mathcal{F}(G_{0}\mathcal{G},\mathcal{E}^2)}
-\frac{(1-i\sqrt{3})}{2}\frac{\mathcal{F}(G_{0}\mathcal{G},\mathcal{E}^2)}{3\left(G_{0} \mathcal{G}\right)^{2}},
\\
\quad
&
\frac{2 \Delta_{0}}{3 G_{0} \mathcal{G}}
-\frac{(1-i\sqrt{3})}{2}\frac{\left(\Delta_{0}^{2}-3\kappa^{2}\right)}{3\mathcal{F}(G_{0}\mathcal{G},\mathcal{E}^2)}
-\frac{(1+i\sqrt{3})}{2}\frac{\mathcal{F}(G_{0}\mathcal{G},\mathcal{E}^2)}{3\left(G_{0} \mathcal{G}\right)^{2}},
\end{align}
where we defined a function
\begin{align}
\mathcal{F}(G_{0}\mathcal{G},|\mathcal{E}|^2)
&:=
2^{-1/3}\left[27 |\mathcal{E}|^{2}\left(G_{0} \mathcal{G}\right)^{4}-2 \Delta_{0}^{3}\left(G_{0} \mathcal{G}\right)^{3}-18 \Delta_{0}\left(G_{0} \mathcal{G}\right)^{3} \kappa^{2}\right. 
\nonumber
\\
&\left.
+\sqrt{
\left(G_{0} \mathcal{G}\right)^{6}\left(4 \left(3\kappa^2-\Delta_{0}^{2}\right)^{3}+\left(27 |\mathcal{E}|^{2}\left(G_{0} \mathcal{G}\right)-2\Delta^{3}_{0}-18\Delta_{0}\kappa^2\right)^2\right)
}\right]^{1 / 3}.
\end{align}
Then, the amplitude $\alpha_{\text{s}}$ becomes the function of the laser power P.
Under $\Delta_{0} > \sqrt{3}\kappa$, all the obtained solutions are real.
\end{appendix}

\end{document}